\documentstyle[12pt,oldlfont]{article}
\newcommand{\simgt}{\rlap{\lower 3.5 pt \hbox{$\mathchar \sim$}} \raise 1pt
 \hbox {$>$}}
\newcommand{\simlt}{\rlap{\lower 3.5 pt \hbox{$\mathchar \sim$}} \raise 1pt
 \hbox {$<$}}
\newcommand{\arcosh}{\mathop{\rm arcosh}\nolimits}
\newcommand{\arsinh}{\mathop{\rm arsinh}\nolimits}
\newcommand{\di}{\mathop{{\rm Li}_2}\nolimits}
\newcommand{\tri}{\mathop{{\rm Li}_3}\nolimits}
\newcommand{\cld}{\mathop{{\rm Cl}_2}\nolimits}
\newcommand{\clt}{\mathop{{\rm Cl}_3}\nolimits}

\catcode`@=11
\newcount\@tempcntc
\def\@citex[#1]#2{\if@filesw\immediate\write\@auxout{\string\citation{#2}}\fi
  \@tempcnta\z@\@tempcntb\m@ne\def\@citea{}\@cite{\@for\@citeb:=#2\do
    {\@ifundefined
       {b@\@citeb}{\@citeo\@tempcntb\m@ne\@citea\def\@citea{,}{\bf ?}\@warning
       {Citation `\@citeb' on page \thepage \space undefined}}%
    {\setbox\z@\hbox{\global\@tempcntc0\csname b@\@citeb\endcsname\relax}%
     \ifnum\@tempcntc=\z@ \@citeo\@tempcntb\m@ne
       \@citea\def\@citea{,}\hbox{\csname b@\@citeb\endcsname}%
     \else
      \advance\@tempcntb\@ne
      \ifnum\@tempcntb=\@tempcntc
      \else\advance\@tempcntb\m@ne\@citeo
      \@tempcnta\@tempcntc\@tempcntb\@tempcntc\fi\fi}}\@citeo}{#1}}
\def\@citeo{\ifnum\@tempcnta>\@tempcntb\else\@citea\def\@citea{,}%
  \ifnum\@tempcnta=\@tempcntb\the\@tempcnta\else
   {\advance\@tempcnta\@ne\ifnum\@tempcnta=\@tempcntb \else \def\@citea{--}\fi
    \advance\@tempcnta\m@ne\the\@tempcnta\@citea\the\@tempcntb}\fi\fi}
\catcode`@=12
\begin{document}
\title{\vskip-3cm{\baselineskip14pt
\centerline{\normalsize DESY 94-083\hfill ISSN 0418-9833}
\centerline{\normalsize May 1994\hfill hep-ph/9405299}}
\vskip1.5cm
Two-Loop ${\cal O}(\alpha_sG_Fm_t^2)$ Corrections to the Fermionic Decay Rates
of the Standard-Model Higgs Boson}
\author{Bernd A. Kniehl\\
II. Institut f\"ur Theoretische Physik, Universit\"at Hamburg\\
Luruper Chaussee 149, 22761 Hamburg, Germany}
\date{}
\maketitle
\begin{abstract}
Low- and intermediate mass Higgs bosons decay preferably into
fermion pairs.
The one-loop electroweak corrections to the respective decay rates
are dominated by a flavour-independent term of ${\cal O}(G_Fm_t^2)$.
We calculate the two-loop gluon correction to this term.
It turns out that this correction screens the leading high-$m_t$ behaviour
of the one-loop result by roughly 10\%.
We also present the two-loop QCD correction to the contribution induced
by a pair of fourth-generation quarks with arbitrary masses.
As expected, the inclusion of the QCD correction considerably
reduces the renormalization-scheme dependence of the prediction.
\end{abstract}

\section{Introduction}

One of the great puzzles of elementary particle physics today
is whether nature makes use of the Higgs mechanism of
spontaneous symmetry breaking to generate the observed particle masses.
The Higgs boson, $H$, is the missing link sought to verify
this concept in the Standard Model.
Many of the properties of the Higgs boson are fixed, e.g.,
its couplings to the gauge bosons,
$g_{VVH}=2^{5/4}G_F^{1/2}M_V^2\ (V=W,Z)$, and fermions,
$g_{f\bar fH}=2^{1/4}G_F^{1/2}m_f$,
and the vacuum expectation value,
$v=2^{-1/4}G_F^{-1/2}\approx246\ {\rm GeV}$.
However, its mass, $M_H$, and its self-couplings, which depend on $M_H$,
are essentially unspecified.

The failure of experiments at LEP~1 and SLC to detect the decay
$Z\rightarrow f\bar f H$ has ruled out the mass range $M_H\le63.8$~GeV at
the 95\% confidence level \cite{sch}.
At the other extreme, unitarity arguments in intermediate-boson scattering
at high energies \cite{dic} and considerations concerning the range of
validity of perturbation theory \cite{vel} establish an upper bound on
$M_H$ at $\left(8\pi\sqrt2/3G_F\right)^{1/2}\approx1$~TeV in a weakly
interacting Standard Model.

The Higgs-boson discovery potential of LEP~1 and SLC is almost exhausted
\cite{gro}.
Prior to the advent of the LHC, the Higgs-boson search will be restricted
to the lower mass range.
With LEP~2 it should be possible to find a Higgs boson with $M_H\le100$~GeV
when high energy and luminosity can be achieved \cite{gkw}.
A possible 4-TeV upgrade of the Tevatron might cover the $M_H$ range up to
120~GeV or so \cite{gun}.
At an $e^+e^-$ linear collider operating at 300~GeV,
50~fb$^{-1}$ luminosity and a $b$-tagging efficiency of 50\%
would be sufficient to detect a Higgs boson with $M_H\le150$~GeV in
the $\mu^+\mu^-b\bar b$ channel \cite{imh}.

Below the onset of the $W^+W^-$ threshold,
the Standard-Model Higgs boson is relatively long-lived,
with $\Gamma_H<100$~MeV, so that, to good approximation,
its production and decay processes may be treated independently.
The low-mass Higgs boson, with $M_H\le M_Z$, decays with more than
99\% probability into a fermion pair \cite{bar}.
With $M_H$ increasing, the $W^+W^-$ mode, with at least one
$W$ boson being off shell, gradually gains importance.
Its branching fraction surpasses that of the $\tau^+\tau^-$ mode at
$M_H\approx115$~GeV and that of the $b\bar b$ mode at $M_H\approx135$~GeV
\cite{bar}.
In the near future, however, Higgs-boson searches will rely mostly on the
$f\bar f$ modes.

Quantum corrections to Higgs-boson phenomenology have received
much attention in the literature; for a review, see Ref.~\cite{bak}.
The experimental relevance of radiative corrections to the $f\bar f$
branching fractions of the Higgs boson has been emphasized recently
in the context of a study dedicated to LEP~2 \cite{gkw}.
Techniques for the measurement of these branching fractions at a
$\sqrt s=500$~GeV $e^+e^-$ linear collider have been elaborated in
Ref.~\cite{hil}.
The QCD corrections to the $H\to q\bar q$ decay rates are most significant
numerically \cite{bra}.
In the approximation $m_q\ll M_H$, they are known to
${\cal O}\left(\alpha_s^2\right)$
\cite{gor}.
The theoretical uncertainty related to the lack of knowledge of the terms of
${\cal O}\left(\alpha_s^2m_q^2/M_H^2\right)$ and
${\cal O}\left(\alpha_s^3\right)$
is presumably small \cite{kat}.
The bulk of the QCD corrections is attributed to the running of $m_q$ up
to scale $M_H$.
In the case of the $b\bar b$ mode,
the QCD correction relative to the Born approximation implemented with
the pole mass ranges between $-53\%$ and $-63\%$ for $M_H$ between 60~GeV
and $2M_W$ \cite{bak}.

The leading high-$M_H$ correction to the $H\to f\bar f$ decay widths is
flavour independent.
The one-loop term, of ${\cal O}\left(G_FM_H^2\right)$, was first obtained by
Veltman \cite{mve};
it is positive and reaches 11\% at $M_H=1$~TeV.
Recently, the two-loop ${\cal O}\left(G_FM_H^2\right)$ term has been found
\cite{dur};
it is negative and exceeds in magnitude the ${\cal O}\left(G_FM_H^2\right)$
term already at $M_H\approx400$~GeV.
The leading contributions due to new heavy fermions are also independent of
the final-state flavour;
at one loop, they are positive and increase quadratically with the
heavy-fermion masses \cite{cha}.
The full one-loop electroweak corrections to the $H\to f\bar f$ decay widths
are now well established \cite{fle,hff}.
They consist of an electromagnetic and a weak part, which are separately
finite and gauge independent.
The electromagnetic part emerges from the one-loop QCD correction \cite{bra}
by substituting $\alpha Q_f^2$ for $\alpha_sC_F$,
where $Q_f$ is the electric charge of $f$ and
$C_F=\left(N_c^2-1\right)/(4N_c)$,  with $N_c=3$.
For $M_H\ll2M_W$, the weak part is well approximated by \cite{hff}
\begin{equation}
\label{weak}
\Delta_{\rm weak}={G_F\over8\pi^2\sqrt2}\left\{{N_c\over3}K_fm_t^2
+M_W^2\left({3\over s_w^2}\ln c_w^2-5\right)
+M_Z^2\left[{1\over2}-3\left(1-4s_w^2|Q_f|\right)^2\right]\right\},
\end{equation}
where $c_w^2=1-s_w^2=M_W^2/M_Z^2$, $K_b=1$, and $K_f=7$ for all other
flavours, except for top.
The $t\bar t$ mode will not be probed experimentally anytime soon
and we shall not be concerned with it in the remainder of this paper.

Throughout this paper, we adopt the so-called modified on-mass-shell
(MOMS) scheme \cite{hol},
which emerges from the ordinary electroweak on-mass-shell scheme \cite{fle,aok}
by eliminating $\alpha$ in favour of $G_F$ by virtue of the relation
$G_F=\left(\pi\alpha/\sqrt2s^2M_W^2\right)(1-\Delta r)^{-1}$ \cite{sir}.
Here, $\Delta r$ embodies the non-photonic correction to the muon decay
width.
In the Born approximation of the MOMS scheme, one has \cite{res}
\begin{equation}
\Gamma_0\left(H\to f\bar f\right)={N_cG_FM_Hm_f^2\over4\pi\sqrt2}
\left(1-{4m_f^2\over M_H^2}\right)^{3/2},
\end{equation}
where $N_c=1$ for lepton flavours,
and the weak correction is implemented by including the overall factor
$(1+\Delta_{\rm weak})$.

Equation~(\ref{weak}) has been obtained by putting $M_H=m_f=0$
($f\ne t$) in the expression for the full one-loop weak correction.
It provides a very good approximation for $f=\tau$ up to
$M_H\approx135$~GeV and for $f=b$ up to $M_H\approx70$~GeV,
the relative deviation from the full weak correction being less than 15\%
in each case.
 From Eq.~(\ref{weak}) it is evident that the dominant effect is due to virtual
top quarks.
In the case $f\ne b$, the $m_t$ dependence is carried solely by the
renormalizations of the wave function and the vacuum expectation value
of the Higgs field and is thus flavour independent.
These corrections are of the same nature as those considered in
Ref.~\cite{cha}.
For $f=b$, there are additional $m_t$ dependent contributions from the
$b\bar bH$ vertex correction and the $b$-quark wave-function renormalization.
Incidentally, they cancel almost completely the universal $m_t$ dependence.
It is amusing to observe that a similar situation has been encountered in the
context of the $Z\to f\bar f$ decays \cite{akh}.
In summary, the universal virtual-top-quark term will constitute the
most important part of the weak one-loop corrections to Higgs-boson
decays in the near future.
In this paper, we shall present the two-loop gluon corrections to this term.

This paper is organized as follows:
In Sect.~2, we shall consider the universal contribution to the
$H\to f\bar f$ decay rates induced by a pair of quarks with arbitrary
masses and evaluate its QCD correction adopting the on-shell definition
of quark mass.
In Sect.~3, we shall repeat this calculation in the modified
minimal-subtraction ($\overline{\rm MS}$) scheme \cite{wab} and
compare the result with the one of Sect.~2 in order to estimate the theoretical
uncertainty related to the arbitrariness of the definition of quark mass.
Section~4 contains the numerical analysis.
Our conclusions are summarized in Sect.~5.

\section{Two-loop results}

In this section, we shall present the QCD correction to the shift in
$\Gamma\left(H\to f\bar f\right)$ induced by a pair of quarks, $(U,D)$,
with arbitrary masses.
For simplicity, we shall assume that $U$ and $D$ do not mix with $f$.
Here, we shall adopt the on-shell definition of quark mass.
As explained in Sect.~1, in the MOMS scheme,
such corrections reside inside the renormalizations of the wave function and
the vacuum expectation value of the Higgs field.
The relevant part of $\Delta_{\rm weak}$ is
\begin{equation}
\label{delta}
\delta=-{\Pi_{WW}(0)\over M_W^2}-\Re e\Pi_{HH}^\prime\left(M_H^2\right),
\end{equation}
where $\Pi_{WW}(s)$ and $\Pi_{HH}(s)$ are the unrenormalized self-energies
of the $W$ and Higgs bosons, respectively,
evaluated at four-momentum squared $s$.
In the following, we shall write down only the $(U,D)$ contributions to the
quantities under consideration.

For completeness, we shall first review the one-loop results.
In dimensional regularization, they read \cite{hzz}
\begin{eqnarray}
\label{hhone}
\Pi_{HH}^0(s)&=&N_c{G_F\over2\pi^2\sqrt2}\sum_{Q=U,D}m_Q^4
\left[\left({s\over2m_Q^2}-3\right)
\left({1\over\epsilon}-\ell_Q\right)
-\left({s\over m_Q^2}-4\right)f\left({s\over4m_Q^2}\right)
\right.\nonumber\\&&\qquad{}+\left.\vphantom{m_Q^2\over\mu^2}
{s\over m_Q^2}-5\right],\\
\label{wwone}
\Pi_{WW}^0(0)&=&N_c{G_FM_W^2\over2\pi^2\sqrt2}\left[
-{1\over2}\sum_{Q=U,D}m_Q^2\left({1\over\epsilon}-\ell_Q+{1\over2}\right)
+{m_U^2m_D^2\over2\left(m_U^2-m_D^2\right)}\ln{m_U^2\over m_D^2}\right],
\end{eqnarray}
where
\begin{equation}
f(r)=\left\{
\begin{array}{l@{\quad:\quad}l}
\displaystyle
\sqrt{1-{1\over r}}\,\arsinh\sqrt{-r} & r<0 \\
\displaystyle
\sqrt{{1\over r}-1}\,\arcsin\sqrt r & 0<r<1 \\
\displaystyle
\sqrt{1-{1\over r}}\,\left(\arcosh\sqrt r-i{\pi\over2}\right) & r>1
\end{array},\right.
\end{equation}
$n=4-2\epsilon$ is the dimensionality of space-time, and
$\ell_Q=\ln\left(m_Q^2/\mu^2\right)$, with $\mu$ being the 't~Hooft mass.
Here and in the following, we suppress terms containing
$\gamma_E-\ln(4\pi)$, where $\gamma_E$ is Euler's constant.
These terms may be retrieved by substituting
$\mu^2\to4\pi e^{-\gamma_E}\mu^2$.
When the on-shell definition of mass is employed, $\mu$ will drop out
in the expressions for physical quantities and so will these terms.
In the $\overline{\rm MS}$ scheme, these terms are subtracted along with
the poles in $\epsilon$.
Inserting Eqs.~(\ref{hhone},\ref{wwone}) in Eq.~(\ref{delta}), one obtains
\cite{hff}
\begin{eqnarray}
\label{delzero}
\delta_0&=&N_c{G_F\over2\pi^2\sqrt2}\left\{\sum_{Q=U,D}m_Q^2
\left[\left(1+{2m_Q^2\over M_H^2}\right)\Re ef\left({M_H^2\over4m_Q^2}\right)
-{1\over4}-{2m_Q^2\over M_H^2}
\right]
\right.\nonumber\\&&\left.
-{m_U^2m_D^2\over2\left(m_U^2-m_D^2\right)}\ln{m_U^2\over m_D^2}\right].
\end{eqnarray}
In the limit $m_U,m_D\gg M_H/2$, this simplifies to \cite{cha}
\begin{equation}
\delta_0=N_c{G_F\over2\pi^2\sqrt2}\left[\sum_{Q=U,D}m_Q^2
\left({7\over12}-{M_H^2\over10m_Q^2}
+{\cal O}\left({M_H^4\over m_Q^4}\right)\right)
-{m_U^2m_D^2\over2\left(m_U^2-m_D^2\right)}\ln{m_U^2\over m_D^2}\right].
\end{equation}
Setting $m_U=m_t$ and $m_D=0$, one recovers the $m_t$-dependent term
of Eq.~(\ref{weak}).
Light quarks, with $m_U,m_D\ll M_H/2$, decouple from $\delta$.

We now proceed to two loops.
The QCD correction to the Higgs-boson self-energy arises from the
Feynman diagrams depicted in Fig.~\ref{one}.
Apart from the two-loop gluon-exchange diagrams [see Fig.~\ref{one}(a)],
one needs to include the diagrams where the one-loop quark mass
counterterm, $\delta m_Q$, is inserted into the vertices and propagators of
the one-loop seed diagram [see Fig.~\ref{one}(b)].

In the on-shell scheme of mass renormalization, $\delta m_Q$ is adjusted
in such a way that the pole of the renormalized propagator appears at the
renormalized mass.
Specifically,
\begin{equation}
{\delta m_Q\over m_Q}=\left.{1\over4m_Q^2}{\rm tr}(\not\!p+m_Q)
\Sigma(p)\right|_{p^2=m_Q^2},
\end{equation}
where
\begin{equation}
\Sigma(p)=i4\pi\alpha_sC_F\left({\mu^2e^{\gamma_E}\over4\pi}\right)^\epsilon
\int{d^nq\over(2\pi)^n}\,{1\over q^2+i\varepsilon}\gamma^\mu
{1\over\not\!q+\not\!p-m_Q+i\varepsilon}\gamma_\mu
\end{equation}
is the quark self-energy due to the exchange of one virtual gluon.
As before, the combination $\gamma_E-\ln(4\pi)$ has been absorbed into a
redefinition of the 't~Hooft mass.
A straightforward calculation yields \cite{hll}
\begin{eqnarray}
\label{osct}
{\delta m_Q\over m_Q}&=&-{\alpha_s\over4\pi}C_F
\left({\mu^2e^{\gamma_E}\over m_Q^2}\right)^\epsilon
{3-2\epsilon\over\epsilon(1-2\epsilon)}\Gamma(1+\epsilon)
\nonumber\\
&=&-{\alpha_s\over4\pi}C_F\left[
{3\over\epsilon}-3\ell_Q+4+\epsilon\left({3\over2}\ell_Q^2-4\ell_Q
+{3\over2}\zeta(2)+8\right)+{\cal O}(\epsilon^2)\right].
\end{eqnarray}

Using a notation consistent with Ref.~\cite{hll}, we find
\begin{equation}
\label{hhtwo}
\Pi_{HH}^1(s)=N_cC_F{G_F\over2\pi^2\sqrt2}\,{\alpha_s\over\pi}\!
\sum_{Q=U,D}\!m_Q^4\left[{9\over2}X_1-\left({s\over4m_Q^2}-3\right)Y_1
-{s\over4m_Q^2}+9\zeta(3)+H_1\left({s\over4m_Q^2}\right)\right],
\end{equation}
where $X_1$ and $Y_1$ are the divergent constants that have been introduced
in connection with the gauge-boson vacuum polarizations at
${\cal O}(\alpha\alpha_s)$ \cite{kni,ks} and $H_1$ is a finite function.
The expressions for $X_1$ and $Y_1$ depend on the regularization scheme.
In dimensional regularization, they read\footnote{As
a consequence of a misprint in Ref.~\cite{djo}, the term of $Y_1$ involving
$\zeta(2)$ occurs in Ref.~\cite{ks} with a wrong prefactor.
Fortunately, this is inconsequential for the physical results of
Ref.~\cite{ks} because the $Y_1$ terms cancel exactly among themselves.}
\begin{eqnarray}
X_1&=&{1\over2\epsilon}-\ell_Q-4\zeta(3)+{55\over12},
\nonumber\\
Y_1&=&{3\over2\epsilon^2}+{1\over\epsilon}\left(-3\ell_Q+{11\over4}\right)
+3\ell_Q^2-{11\over2}\ell_Q+6\zeta(3)+{9\over2}\zeta(2)-{11\over8}.
\end{eqnarray}
For $r<0$,
\begin{eqnarray}
H_1(r)&=&(r-1)\left(2-{1\over r}\right)\left[6\tri\left(r_-^2\right)
-3\tri\left(r_-^4\right)+8f\left(\di\left(r_-^2\right)-\di\left(r_-^4\right)
\right)+4f^2(-3f\right.\nonumber\\
&&{}+\left.\vphantom{\tri\left(r_-^2\right)}g+2h)\right]
+4(r-1)\sqrt{1-{1\over r}}\left[\di\left(r_-^2\right)-\di\left(r_-^4\right)
+f(-3f+2g+4h)\right]\nonumber\\
&&{}+f^2\left(-6r+2+{13\over4r}\right)
+3f\left(-3r+{7\over2}\right)\sqrt{1-{1\over r}}
-{3\zeta(3)\over r}+3\zeta(2)(r-3)+{7\over4},\qquad
\end{eqnarray}
where $\di{}$ and $\tri{}$ are the dilogarithm and trilogarithm \cite{lev},
respectively,
$r_\pm=\sqrt{1-r}\pm\sqrt{-r}\,$, $f=\ln r_+$, $g=\ln(r_+-r_-)$,
and $h=\ln(r_++r_-)$.
A table of handy transformation rules for the analytic continuation in
$r$ is avaliable from Ref.~\cite{zbb}.
For $r>1$, $H_1$ develops an imaginary part, which, by Cutkosky's rule
\cite{cut}, is related to the cuts of the two-loop amplitudes.
This provides the opportunity for a nontrivial check of the calculation,
since this imaginary part is related to the well-known ${\cal O}(\alpha_s)$
correction to $\Gamma\left(H\to q\bar q\right)$.
In fact, for $M_H>2m_q$, one verifies that
\begin{equation}
\Im mH_1\left({M_H^2\over4m_q^2}\right)={\pi\over2}\,{M_H^2\over m_q^2}
\left(1-{4m_q^2\over M_H^2}\right)^{3/2}\delta_{\rm QED},
\end{equation}
where $\delta_{\rm QED}$ is given by Eq.~(3.9) of Ref.~\cite{hff}.

The gluon correction to the $(U,D)$ contribution to $\Pi_{WW}(0)$ may be
found in Ref.~\cite{gam},
where the on-shell definition of quark mass is employed.
The result may be written as
\begin{equation}
\label{wwtwo}
\Pi_{WW}^1(0)={N_cC_F\over4}\,{G_FM_W^2\over2\pi^2\sqrt2}\,{\alpha_s\over\pi}
\left[\sum_{Q=U,D}m_Q^2\left(Y_1-6\zeta(3)-3\zeta(2)+{23\over4}\right)
+F\left(m_U^2,m_D^2\right)\right],
\end{equation}
where
\begin{equation}
\label{fud}
F(u,d)=(u-d)\di\left(1-{d\over u}\right)+{d\over u-d}\ln{u\over d}
\left[u-{3u^2+d^2\over2(u-d)}\ln{u\over d}\right].
\end{equation}
Note that $F(u,d)=F(d,u)$.
 From Eq.~(\ref{fud}), we may read off the properties $F(u,u)=-u$ and
$F(u,0)=\zeta(2)u$.
For $m_D=0$, Eq.~(\ref{wwtwo}) coincides with Eq.~(12) of Ref.~\cite{hll}.

Inserting Eqs.~(\ref{hhtwo},\ref{wwtwo}) into Eq.~(\ref{delta}),
we obtain the general expression for the $(U,D)$ contribution to
$\Gamma\left(H\to f\bar f\right)$ at next-to-leading order,
\begin{equation}
\label{delone}
\delta_1={N_cC_F\over4}\,{G_F\over2\pi^2\sqrt2}\,{\alpha_s\over\pi}
\left[\sum_{Q=U,D}m_Q^2\left(6\zeta(3)+3\zeta(2)-{19\over4}
-\Re eH_1^\prime\left({M_H^2\over4m_Q^2}\right)\right)
-F\left(m_U^2,m_D^2\right)\right].
\end{equation}
For the reader's convenience, we list $\Re eH_1^\prime$ for positive argument.
For $0<r<1$, one has
\begin{eqnarray}
H_1^\prime(r)&=&\left(2-{1\over r^2}\right)\left[
6\clt(2\Phi)-3\clt(4\Phi)+8\Phi\left(\cld(2\Phi)-\cld(4\Phi)\right)
-4\Phi^2(\gamma+2h)\right]\nonumber\\
&&{}+{4\over r}\sqrt{{1\over r}-1}\left[-\cld(2\Phi)+\cld(4\Phi)
+2\Phi(\gamma+2h)\right]+\Phi^2\left(-6+{10\over r}+{5\over4r^2}\right)
\nonumber\\
&&{}-\Phi\left(3+{25\over2r}\right)\sqrt{{1\over r}-1}
+{3\zeta(3)\over r^2}+3\zeta(2)-{9\over2}+{21\over4r}
\end{eqnarray}
and, for $r>1$,
\begin{eqnarray}
\Re eH_1^\prime(r)&=&\left(2-{1\over r^2}\right)\left[
6\tri\left(-\rho_-^2\right)-3\tri\left(\rho_-^4\right)
+8\phi\left(\di\left(-\rho_-^2\right)-\di\left(\rho_-^4\right)\right)
\right.\nonumber\\
&&{}+\left.\vphantom{\di\left(-\rho_-^2\right)}
2(2\phi^2-3\zeta(2))(-3\phi+\gamma+2\chi)\right]
+{2\over r}\sqrt{1-{1\over r}}\left[
2\di\left(-\rho_-^2\right)-2\di\left(\rho_-^4\right)
\right.\nonumber\\
&&{}-\left.9\zeta(2)+2\phi(-3\phi+2\gamma+4\chi)\right]
+\phi^2\left(6-{10\over r}-{5\over4r^2}\right)\nonumber\\
&&{}-\phi\left(3+{25\over2r}\right)\sqrt{1-{1\over r}}
+{3\zeta(3)\over r^2}+3\zeta(2)\left(-2+{5\over r}+{5\over8r^2}\right)
-{9\over2}+{21\over4r},
\end{eqnarray}
where $\cld{}$ and $\clt{}$ are the (generalized) Clausen functions of
second and third order \cite{lev}, respectively,
$\Phi=\arcsin\sqrt r\,$, $\rho_\pm=\sqrt r\pm\sqrt{r-1}\,$,
$\phi=\ln\rho_+$, $\gamma=\ln(\rho_++\rho_-)$, and
$\chi=\ln(\rho_+-\rho_-)$.
It is useful to know the expansions of $H_1^\prime$ appropriate to the
various limiting cases.
They are
\begin{equation}
\label{heavy}
H_1^\prime(r)=6\zeta(3)+3\zeta(2)-{13\over4}+{122\over135}r+{\cal O}(r^2)
\end{equation}
in the heavy-quark limit ($r\ll1$),
\begin{equation}
H_1^\prime(r)=\zeta(2)\left(-12h-6\ln2+{87\over8}\right)-{9\over2}\zeta(3)
+{3\over4}+3\pi\sqrt{1-r}+{\cal O}\Bigl(h(1-r)\Bigr)
\end{equation}
at threshold ($r\simlt1$), and
\begin{equation}
\Re eH_1^\prime(r)=6\gamma^2-3\gamma-6\zeta(2)-{9\over2}+{3\over r}(-6\gamma+1)
+{\cal O}\left({\gamma^2\over r^2}\right)
\end{equation}
in the light-quark limit ($r\gg1$).
 From the above results we can glean the leading QCD correction to $K_f$
for $f\ne b$.
Inserting Eq.~(\ref{heavy}) into Eq.~(\ref{delone}), with $m_U=m_t\gg M_H/2$
and $m_D=0$, and comparing the result with Eq.~(\ref{weak}), one finds the
corrected value,
\begin{eqnarray}
\label{oskf}
K_f&=&7-3\left(\zeta(2)+{3\over2}\right)C_F{\alpha_s\over\pi}\nonumber\\
&=&7-2\left({\pi\over3}+{3\over\pi}\right)\alpha_s\nonumber\\
&\approx&7-4.004\,\alpha_s,
\end{eqnarray}
where terms of ${\cal O}\left(M_H^2/m_t^2\right)$ have been suppressed.
We recover the notion that, in Electroweak Physics, the one-loop
${\cal O}\left(G_Fm_t^2\right)$ terms get screened by their QCD corrections.

\section{Dependence on the quark-mass definition}

So far, we have employed the on-shell definition of quark mass,
i.e., we have evaluated the counterterm diagrams of Fig.~\ref{one}(b)
and similar diagrams for the $W$-boson self-energy using $\delta m_Q$ in the
form specified in Eq.~(\ref{osct}).
This is certainly a reasonable choice.
In the approximation of neglecting the $p^2$ dependence of the imaginary
part of the quark self-energy, the on-shell mass coincides with the real part
of the complex pole position, i.e., with the physical mass,
which is a constant of nature \cite{wil}.
In general, the physical mass is close to what is determined experimentally,
e.g., in quarkonium spectroscopy.
In the case of the top quark, it is approximately the physical mass
that is being extracted at the Tevatron and will be at future $e^+e^-$ linear
colliders, since, in the propagation of the $t$ and $\bar t$ quarks
between the production and decay vertices, configurations near the
mass shell are greatly enhanced kinematically.
As a matter of principle, however, this mass convention is arbitrary,
and one might as well adopt another one.
For, when the perturbation series is summed,
the final result should not depend on the selected scheme.
Yet, this holds no longer true when the perturbation series is truncated.
In general, the finite-order results depend also on the renormalization scales
of the quark masses.
Scheme and typical scale variations may be used to estimate the theoretical
uncertainty due to the unknown higher-order corrections.

In perturbative-QCD calculations, the quark masses are frequently defined
according to the $\overline{\rm MS}$ scheme.
In this case, $\delta m_Q$ collects just the pole in $\epsilon$,
\begin{equation}
{\delta\overline m_Q\over m_Q}=-{\alpha_s\over4\pi}C_F{3\over\epsilon}.
\end{equation}
The relationship between the on-shell mass and the $\overline{\rm MS}$ mass,
$\overline m_Q$, may be read off from Eq.~(\ref{osct}) \cite{tar},
\begin{equation}
\label{msmass}
\overline m_Q=m_Q\left[1+{\alpha_s\over4\pi}C_F(3\ell_Q-4)
+{\cal O}\left(\alpha_s^2\right)\right].
\end{equation}
The mass renormalization scale, $\mu$, must be chosen judiciously according
to the problem at hand so as to minimize higher-order corrections.
The ${\cal O}\left(\alpha_s^2\right)$ term of Eq.~(\ref{msmass}) may be found
in Ref.~\cite{gra}, but we shall not need it here.

In the following, we shall translate the results of Sect.~2 to the
$\overline{\rm MS}$ scheme.
That is, we have to repeat the evaluation of the two-loop counterterm diagrams
using $\delta\overline m_Q$ instead of $\delta m_Q$.
To this end, we may exploit the fact that the corresponding expressions may be
constructed from the one-loop amplitudes by variation, e.g.,
\begin{equation}
\delta\Pi_{HH}^1(s)=\sum_{Q=U,D}\delta m_Q{\partial\over\partial m_Q}
\Pi_{HH}^0(s),
\end{equation}
and similarly for the $\overline{\rm MS}$ scheme.
Consequently, the $\overline{\rm MS}$ version of $\Pi_{HH}^1$ may be obtained
from Eq.~(\ref{hhtwo}) by including the term
\begin{equation}
\Delta\Pi_{HH}^1(s)=\sum_{Q=U,D}\left(\delta\overline m_Q-\delta m_Q\right)
{\partial\over\partial m_Q}\Pi_{HH}^0(s).
\end{equation}
Since $\delta\overline m_Q-\delta m_Q$ is devoid of ultraviolet singularities,
knowledge of the ${\cal O}(\epsilon)$ term of $\Pi_{HH}^0$ is not necessary.
After carrying out this operation in Eqs.~(\ref{hhone},\ref{wwone}), we may
represent the result by assigning shifts to the various items appearing in
Eqs.~(\ref{hhtwo},\ref{wwtwo}),
\begin{eqnarray}
\Delta X_1&=&0,\\
\Delta Y_1&=&{1\over\epsilon}(3\ell_Q-4)-{9\over2}\ell_Q^2
+8\ell_Q-{3\over2}\zeta(2)-8,\\
\label{delhh}
\Delta H_1(r)&=&(3\ell_Q-4)\left[f(2r-5)\sqrt{1-{1\over r}}-2r+{9\over2}
\right],\\
\Delta F\left(m_U^2,m_D^2\right)&=&(3\ell_U-4)m_U^2\left[
{m_D^2\over m_U^2-m_D^2}\left({m_D^2\over m_U^2-m_D^2}\ln{m_U^2\over m_D^2}-1
\right)-{1\over2}\right]+(U\leftrightarrow D).\qquad
\end{eqnarray}
A complimentary set of shifts appropriate to the gauge-boson vacuum
polarizations at arbitrary four-momentum may be found in Ref.~\cite{scheme}.
Similarly to Eq.~(\ref{hhone}), Eq.~(\ref{delhh}) is valid for $r<0$.
To compute the shift of Eq.~(\ref{delone}), we need $\Delta\Re eH_1^\prime$
for $r>0$.
This is
\begin{equation}
\Delta H_1^\prime(r)=(3\ell_Q-4)\left[{\Phi\over\sqrt{1/r-1}}
\left(-2+{1\over r}+{5\over2r^2}\right)-1-{5\over2r}\right]
\end{equation}
for $0<r<1$ and
\begin{equation}
\Delta\Re eH_1^\prime(r)=(3\ell_Q-4)\left[{\phi\over\sqrt{1-1/r}}
\left(2-{1\over r}-{5\over2r^2}\right)-1-{5\over2r}\right]
\end{equation}
for $r>1$.

In the case of the $(t,b)$ contribution to the fermionic decay widths
of an intermediate-mass Higgs boson, we may set
$m_U=m_t\gg M_H/2$ and $m_D=0$.
Then, the shift in Eq.~(\ref{delone}) becomes
\begin{equation}
\Delta\delta_1={N_cC_F\over4}\,{G_Fm_t^2\over2\pi^2\sqrt2}\,{\alpha_s\over\pi}
(3\ell_t-4)\left(-{7\over6}+{\cal O}\left({m_H^4\over m_t^4}\right)\right).
\end{equation}
Thus, the $\overline{\rm MS}$ version of Eq.~(\ref{oskf}) is given by
\begin{equation}
\overline K_f=7+\left(-3\zeta(2)+{19\over2}-{21\over2}\ell_t\right)
C_F{\alpha_s\over\pi},
\end{equation}
where terms of ${\cal O}\left(M_H^2/m_t^2\right)$ have been neglected.
For $\mu=m_t$, this is $7+(2/3)(19/\pi-\pi)\alpha_s\approx7+1.938\,\alpha_s$.
That is, the magnitude of the QCD correction is about half as large as in
the on-shell case and its sign is opposite.

\section{Numerical analysis}

We are now in a position to explore the phenomenological consequences
of our results.
To start with, we specify the input values for our numerical analysis.
We use $M_W=80.24$~GeV \cite{lep}, $M_Z=91.1895$~GeV \cite{blo},
$m_\tau=1.777$~GeV \cite{wei}, $m_b=4.72$~GeV \cite{dom}, and
$m_t=(174\pm16)$~GeV \cite{cdf}.
We parameterize the hadronic contribution to the photon vacuum polarization
according to Ref.~\cite{jeg}, with the updated reference value
$\Delta\alpha_{hadrons}=0.0283$ at $\sqrt s=91.175$~GeV \cite{mar},
and use an equivalent set of effective mass parameters for the light quarks
otherwise \cite{jeg}.
For $\alpha_s(\mu)$, we employ the $\overline{\rm MS}$ formula of
Ref.~\cite{wjm}
and fix the asymptotic scale parameter appropriate to five active flavours,
$\Lambda_{\overline{\rm MS}}^{(5)}$, by requiring that $\alpha_s(M_Z)=0.124$
\cite{blo}.
Unless stated otherwise, we shall choose $\mu=m_t$ for $(t,b)$ contributions
and $\mu=M_H$ else.
All other input parameters are adopted from Ref.~\cite{pdg}.

In Fig.~\ref{two}, we show versus $M_H$ the radiative corrections to the
leptonic decay widths of the Higgs boson originating from quark loops
(dotted lines) and one-gluon exchanges within these loops for
$m_t=(174\pm16)$~GeV.
These corrections are mainly due to the $(t,b)$ doublet and do not depend on
the flavour of the final-state fermions.
As we have observed already in Sect.~2, the QCD correction reduces the
one-loop quark contribution.
For $m_t=176$~GeV, the screening effect ranges between 6.5\% and 10.4\% for
$M_H$ between 60 and 200~GeV.

The full electroweak correction does depend on the produced flavour,
as may be seen already from the approximation of Eq.~(\ref{weak}).
Figure~\ref{two} presents the $M_H$ dependence of the full electroweak
one-loop \cite{hff} (dotted lines) plus QCD two-loop correction to
$\Gamma(H\to\tau^+\tau^-)$ for $m_t=(174\pm16)$~GeV.
Since the QED correction depends logarithmically on $M_H$, the slopes
of the curves are steeper than in Fig.~\ref{two}.
At one loop, there is a large cancellation between the bosonic and fermionic
contributions \cite{hff}.
For $m_t=174$~GeV, their sum is in fact negative and relatively small,
so that the QCD correction enhances significantly the size of the total
correction, by 50.0\% (9.7\%) at $M_H=60$~GeV (150~GeV).

Figures~\ref{two},\ref{three} refer to the on-shell definition of quark mass.
In Figs.~\ref{four},\ref{five}, we study how the radiative corrections
to the leptonic decay widths of the Higgs boson are affected when the
quark mass renormalization is converted from the on-shell scheme to the
$\overline{\rm MS}$ scheme.
For a meaningful study of the scale dependence, it is necessary to distinguish
between the renormalization scales of the quark mass and the strong coupling
constant, $\mu_m$ and $\mu_c$, respectively.
The scale $\mu$ that occurs explicitly in the above formulae must be
identified with $\mu_m$.
Since the quark contributions scale like $m_q^2$
[see Eqs.~(\ref{delzero},\ref{delone})], we may restrict our considerations
to the top quark and neglect the masses of the other quarks.
Figure~\ref{four} visualizes the one- and two-loop evaluations in the on-shell
and $\overline{\rm MS}$ schemes as a function of the top-quark on-shell mass,
$m_t$, for $M_H=100$~GeV.
The solid and dashed lines represent the one- and two-loop on-shell results,
respectively.
The dot-dashed line corresponds to the one-loop evaluation with $m_t$
replaced by $\overline m_t$ as given by Eq.~(\ref{msmass}) with $\mu_m=m_t$.
Inclusion of $\delta_1+\Delta\delta_1$, evaluated with $\overline m_t$,
then leads to the dotted line.
The difference between the dashed and the dotted lines is hardly visible
on the plot, which indicates that the scheme dependence at two loops and
beyond is negligibly small.

In addition to the trivial $\mu_c$ dependence via $\alpha_s$,
which is present also in the on-shell scheme, the $\overline{\rm MS}$
analysis depends on the mass renormalization scale, $\mu_m$, too.
A plausible choice is $\mu_m=m_t$, since this eliminates the terms
proportional to $\ell_t$.
By the same token, it seems unnatural to choose $\mu_m$ very different from
$m_t$, since this renders $\ell_t$ artificially large.
In Fig.~\ref{five}, we analyze the scale dependence of the two-loop
$\overline{\rm MS}$ result for $M_H=100$~GeV and $m_t=174$~GeV.
As usual, we introduce a dimensionless scale parameter, $\xi$,
which we vary from 1/4 to 4.
The dotted line corresponds to the choice $\mu_c=\xi m_t$, $\mu_m=m_t$,
which could be studied also in the on-shell scheme.
Since the QCD corrections are known only to leading order, the $\xi$
dependence is monotonic.
It is expected to flatten when three-loop QCD corrections will be taken into
account.
The dashed line represents the evaluation with $\mu_c=m_t$, $\mu_m=\xi m_t$.
Here, we observe a stabilization already at two loops.
This is due to a cancellation between the $\mu_m$ dependence induced
in the one-loop result via $\overline m_t$ and the one carried
by the genuine two-loop term.
There is no obvious reason to choose $\mu_m$ different from $\mu_c$.
When we identify the two scales and set $\mu_c=\mu_m=\xi m_t$,
we obtain the solid curve,
which assumes its maximum very close to one, at $\xi=0.958$.
This choice may thus be advocated by appealing to the principle of minimal
sensitivity \cite{ste}.

\section{Conclusions}

In the Standard Model, low- and intermediate-mass Higgs bosons decay preferably
into fermion pairs, and the one-loop electroweak corrections to the respective
decay rates are dominated by terms of ${\cal O}\left(G_Fm_t^2\right)$.
The leading two-loop corrections to these terms are of
${\cal O}\left(\alpha_sG_Fm_t^2\right)$ and ${\cal O}\left(G_F^2m_t^4\right)$.
We have calculated the two-loop gluon correction to the shift in the fermionic
decay rates of the Higgs boson induced by pairs of virtual quarks with
arbitrary masses.
This correction is of ${\cal O}\left(\alpha_sG_Fm_Q^2\right)$,
where $m_Q$ is the mass of the heaviest quark,
and does not depend on the produced fermion flavour.
In the case of lepton-pair production, this is the only QCD correction
up to two loops.
In the case of the hadronic decays of the Higgs boson,
this is the only source of ${\cal O}\left(\alpha_sG_Fm_t^2\right)$ corrections.
A special situation arises in the case of $b\bar b$ final states.
Here, the universal ${\cal O}\left(G_Fm_t^2\right)$ term of the electroweak
one-loop correction is almost entirely cancelled by similar contributions
to the $b\bar bH$ vertex and the $b$-quark wave function induced by the
charged current.
The two-loop ${\cal O}\left(\alpha_sG_Fm_t^2\right)$ corrections to the
latter contributions are currently under study \cite{spi}.

For $M_H=60$~GeV (200~GeV), QCD effects screen the positive universal
${\cal O}\left(G_Fm_t^2\right)$ term by 6.5\% (10.4\%),
provided the top quark mass is renormalized according to the on-shell scheme.
In the $\overline{\rm MS}$ scheme, the QCD correction has a different sign
and its magnitude is roughly half as large.
However, the sum of the one- and two-loop corrections is very insensitive
to the choice of mass renormalization scheme.
In the $\overline{\rm MS}$ evaluation with a common renormalization scale,
$\mu$, for the strong coupling constant and the top-quark mass,
the total correction is stable under variations of $\mu$ in the vicinity of
$\mu=m_t$.
This indicates that the effect of QCD corrections beyond two loops is likely
to be insignificant.

\bigskip
\centerline{\bf ACKNOWLEDGMENTS}
\smallskip
The author is indebted to Abdel Djouadi, Paolo Gambino, Kurt Riesselmann,
and Michael Spira for very useful discussions.
He would like to express his gratitude to the Physics Department of UW-Madison
for supporting a visit, during which part of this work was carried out,
and for the great hospitality extended to him.

\begin{figure}[p]

\vskip-6cm

\centerline{\bf FIGURE CAPTIONS}

\caption{\label{one}Feynman diagrams pertinent to the QCD corrections to the
Higgs-boson self-energy: (a) gluon exchanges and (b) counterterms.}

\caption{\label{two}${\cal O}(\alpha)$ and ${\cal O}(\alpha\alpha_s)$
radiative corrections to $\Gamma\left(H\to\ell^+\ell^-\right)$, with
$\ell=e,\mu,\tau$, due to virtual quarks as a function of $M_H$ for
$m_t=(174\pm16)$~GeV.}

\caption{\label{three}Full ${\cal O}(\alpha)$ and ${\cal O}(\alpha\alpha_s)$
radiative corrections to $\Gamma\left(H\to\tau^+\tau^-\right)$ as a function
of $M_H$ for $m_t=(174\pm16)$~GeV.}

\caption{\label{four}${\cal O}(\alpha)$ and ${\cal O}(\alpha\alpha_s)$
radiative corrections to $\Gamma\left(H\to\ell^+\ell^-\right)$, with
$\ell=e,\mu,\tau$, due to virtual quarks as a function of $m_t$ for
$M_H=100$~GeV.
The evaluations using the on-shell and $\overline{\rm MS}$ definitions of the
top-quark mass are compared.
All other quark masses are set to zero.}

\caption{\label{five}${\cal O}(\alpha\alpha_s)$ radiative corrections to
$\Gamma\left(H\to\ell^+\ell^-\right)$, with $\ell=e,\mu,\tau$, for
$M_H=100$~GeV and $m_t=174$~GeV evaluated in the $\overline{\rm MS}$ scheme
with $\mu_c=\xi m_t$, $\mu_m=m_t$ (dotted line), $\mu_c=m_t$, $\mu_m=\xi m_t$
(dashed line), and $\mu_c=\mu_m=\xi m_t$ (solid line).
All other quark masses are set to zero.}
\end{figure}

\end{document}